\documentclass[letterpaper,twocolumn,aps,amsmath,amssymb,floatfix]{revtex4-1}
\usepackage{amsmath,amssymb,amsfonts,mathrsfs}
\usepackage{color}
\usepackage{epsfig}
\usepackage{psfrag}
\usepackage{graphicx}
\usepackage{color}
\usepackage{bm}

\begin{document}

\title{Equilibration of Quasi-One-Dimensional Fermi Gases}

\author{Wade DeGottardi$^{1,2}$ and K. A. Matveev$^3$}

\affiliation{\small{$^1$Institute for the Research in Electronics and Applied Physics, University of Maryland, College Park, Maryland 20742, USA}}

\affiliation{\small{$^2$Joint Quantum Institute, NIST/University of Maryland, College Park, Maryland, 20742, USA}}

\affiliation{\small{$^3$Materials Science Division, Argonne National Laboratory, Argonne, Illinois 60439, USA }}
\date{\today}

\begin{abstract}
One-dimensional systems often possess multiple channels or bands arising from the excitation of transverse degrees of freedom. In the present work, we study the specific processes that dominate the equilibration of multi-channel Fermi gases at low temperatures. Focusing on the case of two channels, we perform an analysis of the relaxation properties of these systems by studying the spectrum and eigenmodes of the linearized collision integral. As an application of this analysis, a detailed calculation of the bulk viscosity is presented. The dominant scattering processes obey an unexpected conservation law which is likely to affect the hydrodynamic behavior of these systems.
\end{abstract}

\maketitle

\section{Introduction}

One-dimensional (1D) Fermi systems exhibit a host of exotic phenomena which signal the breakdown of Landau's Fermi liquid theory~\cite{haldane,deshpande,giamarchi}. Ultimately, this failure can be traced to the combined effects of the Pauli exclusion principle and the restricted nature of motion in one dimension. These effects play an important role in constraining the type of collisions that allow these systems to relax to equilibrium~\cite{beyond,daley}. For a single channel system, two particle scattering cannot alter the fermion occupation numbers---a consequence of energy and momentum conservation in one dimension. Instead, relaxation must occur via three-particle collisions~\cite{beyond}. These processes have rates that scale as $1/\tau \propto T^7$, and thus are strongly suppressed at low temperatures~\cite{mobile,gutman,beyond}. This has important implications for the thermal and mechanical responses of these systems.

Most examples of 1D Fermi systems are \emph{quasi}-1D in certain regimes, i.e., these systems possess additional channels (or bands) which arise from the excitation of transverse degrees of freedom. Prominent examples include quantum wires~\cite{vanWees,pepper} and cold atomic Fermi gases~\cite{kinoshita,moritz}. The presence of multiple bands has a profound effect on the manner in which 1D systems equilibrate. In contrast to the single channel case, the occupation of the states in a multi-channel system can be altered by two-particle scattering processes~\cite{us}. Since these processes occur at a rate $1/\tau \propto T$, they lead to more rapid equilibration than the three-particle collisions discussed above. Thus, at sufficiently low temperatures, multi-channel electron systems will relax more quickly than single channel systems.

The number of occupied channels depends on the fermion density and in many quasi-1D systems is thus readily tunable. In quantum wires, the density can be controlled by the gate voltage~\cite{vanWees,pepper}. In cold atomic systems, the density can be altered by changing the number of atoms loaded into the trap~\cite{kinoshita,moritz}. Therefore, the effects associated with the occupation of multiple channels are of particular experimental relevance. The dramatic difference between the equilibration properties of one- and multi-channel systems should be reflected in their respective transport coefficients. A transport coefficient of particular interest in one dimension is the bulk viscosity $\zeta$, which describes the viscous forces that arise when a fluid expands or contracts~\cite{landau-fluids,pustilnik}. In quantum wires, the bulk viscosity determines the amount of resistance that arises from spatial variations of the electron density~\cite{hydro,transport}. In cold atomic systems, $\zeta$ determines the damping rate of a breathing mode of the atomic gas~\cite{vogt}.

Focusing on the two-channel case, we present a detailed study of the equilibration properties of spinless multi-channel Fermi gases. We diagonalize the linearized collision integral describing the two-body collisions that dominate the relaxation properties of these systems at low temperatures. As an application of our analysis of the collision integral, we present a careful calculation of the bulk viscosity $\zeta$. In earlier work, $\zeta$ was estimated using the relaxation time approximation, a phenomenological approach in which all the non-equilibrium states are assumed to relax at the same rate~\cite{us}. In contrast, the approach presented here accounts for the microscopic details of the relaxation processes, thus enabling us to determine how $\zeta$ depends on the properties of the system, such as the Fermi velocities of the channels.

An important feature of the system, revealed in the relaxation spectrum, is an unexpected conservation law that is respected by the dominant scattering processes. A full hydrodynamic description of the system would require taking this conservation law into account. Given that conservation laws underly hydrodynamics, this additional conserved quantity is likely to give rise to behavior which differs markedly from that predicted by standard hydrodynamics~\cite{landau-fluids}. This conservation law is violated by sub-leading scattering processes and in systems that do not possess Galilean invariance. The implications of this violation on the non-equilibrium behavior of such systems are discussed.

The organization of this paper is the following. In Sec.~\ref{sec:processes}, we identify the two-body processes that dominate the relaxation properties of multi-channel 1D Fermi gases. In Sec.~\ref{sec:lifetimes}, the quasiparticle lifetimes are calculated for $T \to 0$. The linearized collision integral is analyzed and the relaxation spectrum of the system is obtained in Sec.~\ref{sec:spectrum}. The calculation of the bulk viscosity is presented in Sec.~\ref{sec:viscosity}. In Sec.~\ref{sec:Q}, the unexpected conservation law described above is addressed. Finally, in Sec.~\ref{sec:conclusion}, we discuss our results.

\begin{figure}
\begin{center}
\includegraphics[width = 8cm]{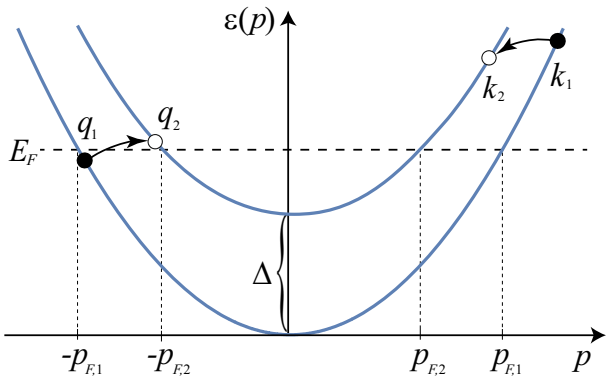}
\label{fig:spectrum}
\caption{The scattering processes that dominate the relaxation properties of a two-channel Fermi gas. In the particular process shown, right- and left-moving fermions in band 1 scatter into band 2. }
\label{fig:2channelscattering}
\end{center}
\end{figure}

\section{Effective 1D Hamiltonian and Dominant Relaxation Processes}
\label{sec:processes}

Turning to the system of interest, we consider spinless fermions with an effective 1D dispersion
\begin{equation}
\varepsilon_i(p) = \frac{p^2}{2m} + \Delta_i,
\label{eq:quad}
\end{equation}
where $\Delta_i$ is the energy cost associated with occupation of the $p=0$ state of the $i^{\textrm{th}}$ band. The fermion mass is denoted by $m$. For a Fermi energy $E_F$ in the range $\Delta_l < E_F < \Delta_{l+1}$ with the $\{\Delta_i\}$ monotonically increasing in $i$, there are $l$ occupied bands at $T = 0$. The $i^\textrm{th}$ band is filled to Fermi points $\pm p_{F,i}$, where $p_{F,i} = \sqrt{2 m \left( \mu - \Delta_i \right)}$, and is characterized by a Fermi velocity $v_{i} = p_{F,i} / m$. As long as
\begin{equation}
T \ll |E_F - \Delta_i|
\label{eq:lowT}
\end{equation}
for all $i$, the $l$ lowest bands are nearly degenerate and all others are effectively unoccupied.

At low temperatures, typical excitations involve states near the Fermi points. It will thus be convenient to redefine the zero of energy to be at $E_F$. Similarly, the momentum will be measured from the corresponding Fermi point, with $k_i$ and $q_i$ reserved for right- and left-moving particles, respectively. We define
\begin{eqnarray}
\epsilon_{k_i} &=& \varepsilon_i(k_i + p_{F,i}) - E_F, \label{eq:epsilon1}\\
\epsilon_{q_i} &=& \varepsilon_i(q_i - p_{F,i}) - E_F,
\label{eq:epsilon2}
\end{eqnarray}
which take the form
\begin{eqnarray}
\epsilon_{k_i} &=& v_i k_i + k_i^2/2m, \label{eq:epsilon3A}\\
\epsilon_{q_i} &=& -v_i q_i + q_i^2/2m. \label{eq:epsilon3B}
\label{eq:epsilon4}
\end{eqnarray}
In the temperature regime~(\ref{eq:lowT}), typical excitations are well described by the linear term.

The two-body interaction between fermions is given by
\begin{eqnarray}
\hat{V}_{\textrm{3D}} &=& \frac{1}{2} \int \! dx \, dx' \, d \boldsymbol{\rho} \, d \boldsymbol{\rho}' \, V_{\textrm{3D}}(x-x';\boldsymbol{\rho},\boldsymbol{\rho}') \nonumber \\
 && \times \Psi^\dagger(x,\boldsymbol{\rho}) \Psi^\dagger(x',\boldsymbol{\rho}')  \Psi(x',\boldsymbol{\rho}') \Psi(x,\boldsymbol{\rho}),
\label{eq:V3D}
\end{eqnarray}
where $x$ and $x'$ are coordinates along the 1D system, $\boldsymbol{\rho}$ and $\boldsymbol{\rho}'$ are vectors in the two transverse dimensions, and $V_{\textrm{3D}}$ is the three-dimensional microscopic interaction between fermions. The operator $\Psi(x,\boldsymbol{\rho})$ annihilates a fermion at the position $(x,\boldsymbol{\rho})$.

We will now obtain an effective 1D description of the interactions, i.e., one that does not involve the transverse coordinates. To accomplish this, we write the fermion operators as
\begin{eqnarray}
\nonumber
\Psi(x,\boldsymbol{\rho}) = && \frac{1}{\sqrt{L}} \sum_{i} \varphi_{i}(\boldsymbol{\rho}) \Bigg( \sum_{k_i>-p_{F,i}}  e^{i \left( p_{F,i} + k_i \right) x} a_{k_i} \\&&  + \sum_{q_i<p_{F,i}} e^{i \left(- p_{F,i} + q_i  \right) x} a_{q_i} \Bigg),\label{eq:expansion}
\end{eqnarray}
where $L$ is the length of the system. The $\varphi_i(\boldsymbol{\rho})$ are the wave functions associated with the various channels and the operators $a_{k_i}$ and $a_{q_i}$ are the $i^{\textrm{th}}$-channel fermion annihilation operators for right- and left-moving fermions, respectively. Substituting Eq.~(\ref{eq:expansion}) into Eq.~(\ref{eq:V3D}) and integrating over $\boldsymbol{\rho}$, $\boldsymbol{\rho}'$, $x$, and $x'$, we obtain an effective 1D operator. In the low temperature regime (\ref{eq:lowT}), the dominant interaction terms are given by
\begin{equation}
\hat{V}_{\textrm{1D}}= \frac{1}{L} \sum_{i,j } \sum_{k_i q_i k_j q_j} \lambda_{ij}^{\phantom\dagger} a_{k_j}^\dagger a_{q_j}^\dagger a_{q_i}^{\phantom\dagger} a_{k_i}^{\phantom\dagger} \delta_{k_i + q_i, k_j + q_j}
\label{eq:V1D}
\end{equation}
where
\begin{eqnarray}
\lambda_{ij} = \int dy \, d \boldsymbol{\rho} \, d \boldsymbol{\rho}' \varphi_j^\ast(\boldsymbol{\rho}) \varphi_j^\ast(\boldsymbol{\rho}') \varphi_i^{\phantom\dagger}(\boldsymbol{\rho}) \varphi_i^{\phantom\dagger} (\boldsymbol{\rho}') \, \, \, \, \, \, \nonumber \\
\times \label{eq:Veff} \left( e^{i(p_{F,i}-p_{F,j})y} - e^{i(p_{F,i}+p_{F,j})y} \right) V_{\textrm{3D}}(y;\boldsymbol{\rho},\boldsymbol{\rho}').
\label{eq:lambda}
\end{eqnarray}
We have assumed that the magnitudes of the momenta $k_i$ and $q_i$ are negligible compared to $p_{F,i}$. We have not included terms in $\hat{V}_{\textrm{1D}}$ that can arise only for special values of $p_{F,i}$, see Ref.~\cite{lunde-resonance}.

The operator $\hat{V}_\textrm{1D}$ provides an effective 1D description of the dominant interactions in a multi-channel Fermi gas at low temperatures. To first order, this perturbation generates scattering processes of the form
\begin{equation}
i + i \rightarrow j + j
\label{eq:rxn}
\end{equation}
in which a right- and a left-moving fermion in band $i$ scatter into band $j$. Conservation laws forbid two-particle intra-channel processes, i.e., those for which $i = j$~\cite{beyond}. In contrast, inter-channel scattering, such as the process shown in Fig.~\ref{fig:2channelscattering}, can occur near the Fermi energy and still satisfy energy and momentum conservation at low temperatures~\cite{us}.

\section{Quasiparticle Lifetimes}
\label{sec:lifetimes}

As a preliminary step in characterizing the effect of inter-channel scattering processes, we calculate the lifetimes of quasiparticles for $T \to 0$. We focus on the behavior of a system with two channels.

The rate at which a fermion in band 1 in state $k_1$ scatters is given by Fermi's golden rule,
\begin{eqnarray}
\frac{1}{\tau_1(k_1)} &=& \frac{2 \pi | \lambda |^2}{\hbar L^2} \sum_{k_2 q_1 q_2} \delta_{k_1+q_1,k_2+q_2} \delta(\epsilon_{k_1} + \epsilon_{q_1} - \epsilon_{k_2} - \epsilon_{q_2})  \nonumber \\ && \times  f_{q_1} \left( 1 - f_{k_2} \right)  \left( 1 - f_{q_2} \right),
\label{eq:tauk1}
\end{eqnarray}
where $\lambda = \lambda_{12}$ is defined in Eq.~(\ref{eq:lambda}) and $\hbar$ is Planck's constant. The functions $f_{k_i}$ and $f_{q_i}$ denote the Fermi-Dirac distributions
\begin{equation}
f_{k_i} = \frac{1}{e^{\epsilon_{k_i}/T} + 1},\quad \quad\ f_{q_i} = \frac{1}{e^{\epsilon_{q_i}/T}+1},
\label{eq:fd1}
\end{equation}
for right- and left-moving fermions, respectively. For energies much less than $E_F$, the spectrum (\ref{eq:epsilon3A}) and (\ref{eq:epsilon3B}) is effectively linear and described by
\begin{equation}
\epsilon_{k_i} = v_i k_i, \quad \quad \epsilon_{q_i} = -v_i q_i.
\label{eq:linearspec}
\end{equation}
In order to understand the behavior of the quasiparticle decay rates, we express conservation of momentum in the form
\begin{equation}
k_2 = k_1 + q_1 - q_2.
\label{eq:consmom1}
\end{equation}
Equation~(\ref{eq:consmom1}) can then be used to eliminate $k_2$ from the equation expressing conservation of energy, yielding
\begin{equation}
\left( v_1 + v_2 \right) q_1 - 2 v_2 q_2 = \left(v_1 - v_2\right) k_1.
\label{eq:conseng1}
\end{equation}
Noting that $v_1 > v_2$, it is clear that there are $q_1 > 0$, $q_2 < 0$ and $k_2 > 0$ that satisfy Eqs.~(\ref{eq:consmom1}) and (\ref{eq:conseng1}). This case is depicted in Fig. 1. Thus, even at $T = 0$, the right-hand side of Eq.~(\ref{eq:tauk1}) remains non-zero for these momenta and therefore $\tau_1(k_1)$ is finite. For $T = 0$, an explicit calculation of Eq.~(\ref{eq:tauk1}) gives
\begin{equation}
\frac{1}{\tau_1(k_1)} = \frac{|\lambda|^2 (v_1 - v_2) k_1}{4 \pi \hbar^3 v_2 (v_1+v_2)}.
\label{eq:tauk1form}
\end{equation}
This result remains valid for nonzero temperatures as long as $(v_1-v_2) k_1 \gg T$.

A similar calculation can be performed for $1/\tau_2(k_2)$, the rate of decay of quasiparticles in channel 2. An important element of the above analysis for $\tau_1(k_1) $ was that the fermion was on the faster branch (i.e. $v_1 > v_2$). We find that for $T = 0$ no decay channels exist for a fermion at $k_2$. Thus, $1/\tau_2(k_2)$ will vanish as $T \rightarrow 0$. In the limit $v_2^2 \left( v_1 - v_2 \right)^2 k_1 / v_1^3 \gg T$, the analog of Eq.~(\ref{eq:tauk1}) for $1/\tau_2(k_2)$ can be evaluated to give
\begin{equation}
\frac{1}{\tau_2 (k_2)} = \frac{ | \lambda |^2 T}{2 \hbar^3 v_1 (v_1+v_2) \sin \left( \frac{2 \pi v_2 }{v_1+v_2}\right)} e^{- \frac{v_1-v_2}{v_1 + v_2} \frac{v_2 k_2}{T}}.
\label{eq:tauk2form}
\end{equation}
As anticipated, the rate vanishes for $T = 0$.

In summary, we have found that the behavior of the quasiparticle decay rates differs dramatically for the two channels. For the faster channel (channel 1), the decay rates remain non-zero at zero temperature. In contrast, quasiparticles in channel 2 exhibit decay rates that are suppressed exponentially at low temperatures~\cite{footnote1}. The above calculations are readily
repeated for holes. Due to the particle-hole symmetry of the linearized spectrum, identical behavior is found.

\section{Spectrum of Linearized Collision Integral}
\label{sec:spectrum}

Although the lifetimes of highly excited quasiparticles obtained in the previous section gave us important information about inter-channel relaxation, these lifetimes are not directly relevant to the calculation of transport coefficients. Rather, the non-equilibrium response of a system is dominated by the relaxation of quasiparticles with energies on the order of $T$. Furthermore, the probability that a given state is occupied must be taken into account.

A calculation of the transport coefficients requires the study of the time evolution of the distribution functions $n_{k_i}$ and $n_{q_i}$. Here we focus on the two-channel case with $i = 1,2$. The scattering generated by $\hat{V}_{\textrm{1D}}$ can be calculated using Fermi's golden rule. To lowest order in the perturbation (\ref{eq:V1D}), we find
\begin{eqnarray}
\dot{n}_{k_1} = &-& \frac{2\pi| \lambda |^2 }{\hbar L^2} \sum_{k_2 q_1  q_2} \delta_{k_1+q_1,k_2+q_2} \nonumber \\
&\times& \delta \left(  \epsilon_{k_1} + \epsilon_{q_1} - \epsilon_{k_2} - \epsilon_{q_2} \right) \nonumber \\
&\times&  \bigg[ n_{k_1} n_{q_1} \left( 1 - n_{k_2} \right) \left( 1 - n_{q_2} \right) \nonumber \\
& & \, \, \, \, \,   -  n_{k_2} n_{q_2} \left( 1- n_{k_1} \right) \left( 1 - n_{q_1} \right) \bigg],
\label{eq:fermi}
\end{eqnarray}
where $\dot{n}_{k_i}$ indicates the time derivative of $n_{k_i}$. Similar expressions are obtained for $\dot{n}_{q_1}$, $\dot{n}_{k_2}$, and $\dot{n}_{q_2}$.

The collision integral~(\ref{eq:fermi}) is non-linear and thus resists straightforward analysis. However, a full analysis of the collision integral is not always necessary. For example, transport coefficients describe a system that is only slightly perturbed from equilibrium. In this regime, the collision integral can be linearized in parameters that describe the deviation of a distribution function from equilibrium, thus dramatically simplifying the required analysis.

\subsection{Linearized Collision Integral}

The linearized collision integral provides a description of the system near equilibrium. To obtain it, we expand $n_{k_i}$ around the equilibrium distribution function (\ref{eq:fd1}) as follows
\begin{equation}
n_{k_i} = f_{k_i} + g_{k_i} x_{k_i},
\label{eq:nx}
\end{equation}
where $x_{k_i} = 0$ corresponds to equilibrium and
\begin{equation}
g_{k_i} = \sqrt{ f_{k_i} \left( 1 - f_{k_i} \right)} = \frac{1}{2 \cosh \left(\frac{v_i k_i}{2T} \right)}.
\label{eq:gki}
\end{equation}
(The last equality in Eq.~(\ref{eq:gki}) assumes a linear spectrum (\ref{eq:linearspec}).)
A similar expansion is obtained for the $n_{q_i}$ in terms of $x_{q_i}$. The specific choice of $g_{k_i}$ ensures that the kernel of the linearized collision integral is symmetric, see below. Employing Eq.~(\ref{eq:nx}), Eq.~(\ref{eq:fermi}) can be linearized in the $x_{k_i}$ and $x_{q_i}$ to give
\begin{eqnarray}
 \dot{x}_{k_1} = &-& \frac{2 \pi |\lambda|^2}{\hbar L^2} \sum_{ k_2 q_1 q_2} \delta_{k_1+q_1,k_2+q_2} \nonumber \\
 &\times& \delta \left( \epsilon_{k_1} + \epsilon_{q_1} - \epsilon_{k_2} - \epsilon_{q_2} \right) g_{q_1} g_{k_2} g_{q_2}  \nonumber \\
 &\times& \bigg( \frac{x_{k_1}}{g_{k_1}} + \frac{x_{q_1}}{g_{q_1}} - \frac{x_{k_2}}{g_{k_2}} - \frac{x_{q_2}}{g_{q_2}} \bigg).
  \label{eq:fermi2}
\end{eqnarray}
Analogous expressions are obtained for $\dot{x}_{q_1}$, $\dot{x}_{k_2}$ and $\dot{x}_{q_2}$.

The linearized collision integral can be expressed in the form
\begin{eqnarray}
\nonumber
\dot{x}_{k_1} &=& - \frac{1}{\tau_1(k_1)} x_{k_1} - \int dk_2 \,  \Gamma_{12}^{++}(k_1,k_2) x_{k_2} \\
&& - \int \! dq_1 \Gamma_{11}^{+-}(k_1,q_1) x_{q_1} - \int \! dq_2 \Gamma_{12}^{+-}(k_1,q_2) x_{q_2}. \, \, \, \, \, \, \, \, \, \, \, \label{eq:xdotki}
\end{eqnarray}
Each of the terms on the right-hand side of Eq.~(\ref{eq:xdotki}) arises from one of the four terms within pararentheses in Eq.~(\ref{eq:fermi2}). The kernels $\Gamma_{ii'}^{rr'}$ possess indices $r$ and $r'$ indicating right- or left-moving branches as well as band indices $i$ and $i'$. Analogous expressions for $\dot{x}_{k_2}$, $\dot{x}_{q_1}$, and $\dot{x}_{q_2}$ are readily obtained. Explicit expressions for the collision kernels are given in Appendix A.

The action of the linearized collision integral on the full set of $\{x_{k_i},x_{q_i}\}$ can be written compactly as
\begin{equation}
| \dot{x} \rangle = - \hat{\Gamma} | x \rangle,
\label{eq:xdot}
\end{equation}
where the vector $|x\rangle$ is defined by
\begin{equation}
| x \rangle = \left(
                 \begin{array}{c}
                  x_{k_1} \\
                  x_{k_2} \\
                  x_{q_2} \\
                  x_{q_1} \\
                 \end{array}
               \right).
\label{eq:xvector}
\end{equation}
We remind readers that $x_{k_1}$ encodes the distribution function for right-movers in band 1, the variable $x_{k_2}$ encodes the distribution for right-movers in band 2, and so on.

We employ separation of variables to cast Eq.~(\ref{eq:xdot}) as an eigenvalue problem. The solutions have the form
\begin{equation}
| x(t) \rangle = e^{-\gamma_j t } | X_j \rangle, \\
\end{equation}
where $| X_j \rangle$ is independent of time and is the eigenmode of $\hat{\Gamma}$ with eigenvalue $\gamma_j$, i.e.,
\begin{equation}
\hat{\Gamma} | X_j \rangle = \gamma_j | X_j \rangle.
\label{eq:eigprob}
\end{equation}
We have verified that the kernel $\Gamma_{ii'}^{rr'}$ is symmetric under the simultaneous interchange of its indices and continuous variables, for example $\Gamma_{12}^{+-}(k_1,q_2) = \Gamma_{21}^{-+}(q_2,k_1)$. Since the linearized collision integral $\hat{\Gamma}$ is symmetric and real, the $\gamma_j$ are real. Because $\hat{\Gamma}$ has been obtained by expanding around equilibrium, its eigenvalues $\gamma_j$ must be non-negative.

\subsection{Conserved quantities and zero modes}

\label{sec:zero}

The collision integral respects the conservation laws obeyed by the scattering processes encoded in it. It is thus not surprising that these conservation laws are reflected in the spectrum of $\hat{\Gamma}$. As we discuss below, for each conservation law there is an eigenmode of $\hat{\Gamma}$ with an eigenvalue equal to zero. We will refer to such modes as \emph{zero modes}.

These zero modes may be obtained directly from the equilibrium distribution function. We use the fact that for each conservation law there is a parameter appearing in the equilibrium distribution function. For example, in the grand canonical ensemble the number of particles and energy are fixed by the values of the chemical potential $\mu$ and temperature $T$. Conservation of momentum gives rise to an additional constraint. The associated parameter $u$ is the velocity of the center of mass of the gas. The equilibrium distribution functions have the form
\begin{equation}
f_{i}(p) = \frac{1}{e^{\left( \varepsilon_i(p) - u p - \mu \right)/T}+1},
\label{eq:fieq}
\end{equation}
and reduce to $f_{k_i}$ or $f_{q_i}$ given by Eq.~(\ref{eq:fd1}) for $u = 0$ and $\mu = E_F$.

A small deviation from the equilibrium distribution function $f_i(p)$ typically describes a non-equilibrium state which then relaxes back to equilibrium. However, consider a new equilibrium that arises from changing a parameter in the distribution function (\ref{eq:fieq}), say $T$. The new equilibrium distribution, which may be described by $|x\rangle$ using Eqs.~(\ref{eq:nx}) and (\ref{eq:xvector}), does not relax. Thus, $|\dot{x}\rangle$ equals zero and consequently $|x\rangle$ is a zero mode of $\hat{\Gamma}$, see Eq.~(\ref{eq:xdot}).

Given that Eq.~(\ref{eq:fieq}) has three parameters, we can immediately obtain three zero modes employing this procedure. We begin by considering the total number of particles
\begin{equation}
N = N_{1R} + N_{2R} + N_{2L} + N_{1L},
\end{equation}
where $N_{i,R/L}$ is the number of right/left-moving particles in the $i^{\textrm{th}}$ band. We can obtain the zero mode associated with particle number conservation by varying $\mu$ in Eq.~(\ref{eq:fieq}), i.e. taking $\mu \rightarrow \mu + \delta \mu$ and expanding in $\delta \mu$. Then, setting $u = 0$ and $\mu = E_F$ in the resulting expression gives
\begin{equation}
f_i(p) = f_{k_i} + \frac{\delta \mu}{T} g_{k_i}^2.
\end{equation}
Using Eq.~(\ref{eq:nx}), we identify $x_{k_i}$ as $g_{k_i} \delta \mu / T$. Hence,
\begin{equation}
| N \rangle = \left(
                 \begin{array}{c}
                  g_{k_1} \\
                  g_{k_2} \\
                  g_{q_2} \\
                  g_{q_1} \\
                 \end{array}
               \right)
\end{equation}
is a zero mode of $\hat{\Gamma}$. That $|N\rangle$ is indeed a zero mode can be verified by substituting $| N \rangle$ for $|x\rangle$ in Eq.~(\ref{eq:fermi2}). This gives $x_{k_i} = g_{k_i}$ and $x_{q_i} = g_{q_i}$, for which the right-hand side of Eq.~(\ref{eq:fermi2}) clearly vanishes.

Energy conservation and momentum conservation give rise to two additional zero modes. The zero mode associated with energy conservation can be obtained by varying $T$ in Eq.~(\ref{eq:fieq}). Expanding in $\delta T$ and setting $\mu = E_F$ and $u$ to zero gives the zero mode
\begin{equation}
| E \rangle = \left(
\begin{array}{c}
     g_{k_1}  \epsilon_{k_1} \\
     g_{k_2}  \epsilon_{k_2} \\
     g_{q_2}  \epsilon_{q_2} \\
     g_{q_1}  \epsilon_{q_1}  \\
    \end{array}
    \right).
\label{eq:Ezero}
\end{equation}
Varying $u$ in Eq.~(\ref{eq:fieq}) gives
\begin{equation}
| P \rangle = \left(
                 \begin{array}{c}
                  g_{k_1} k_1 \\
                  g_{k_2} k_2 \\
                  g_{q_2} q_2 \\
                  g_{q_1} q_1 \\
                 \end{array}
               \right).
\label{eq:Pzero}
\end{equation}
Substituting Eqs.~(\ref{eq:Ezero}) and (\ref{eq:Pzero}) into Eq.~(\ref{eq:fermi2}) allows us to verify that $|E\rangle$ and $|P\rangle$ are indeed zero modes.

The conservation of particle number, energy, and momentum is a generic feature of scattering processes. For the specific processes shown in Fig.~1, two additional conserved quantities can be identified, namely,
\begin{equation}
J_1 = N_{1R} - N_{1L},
\end{equation}
and
\begin{equation}
J_2 = N_{2R} - N_{2L}.
\end{equation}
The above analysis can be repeated with these two additional conservation laws. We obtain an equilibrium distribution function similar to Eq.~(\ref{eq:fieq}), but with two additional parameters that fix the values of $J_1$ and $J_2$. The corresponding zero modes are
\begin{equation}
|J_1\rangle =
\left(
  \begin{array}{r}
  g_{k_1}  \\
  0 \, \, \, \, \\
  0 \, \, \, \, \\
  - g_{q_1}
  \end{array}
\right)
\label{eq:J1}
\end{equation}
and
\begin{equation}
|J_2\rangle =
\left(
  \begin{array}{r}
    0  \, \, \, \, \\
     g_{k_2}  \\
   - g_{q_2} \\
    0   \, \, \, \, \\
  \end{array}
\right).
\label{eq:J2}
\end{equation}
Substituting these expressions for $|J_1\rangle$ and $|J_2\rangle$ into Eq.~(\ref{eq:fermi2}) confirms that they are indeed zero modes. For the case of $l$ channels, there will be $l$ conservation laws in addition to the conservation of particle number, energy, and momentum.

\begin{center}
\begin{figure*}
\includegraphics[width = 17cm]{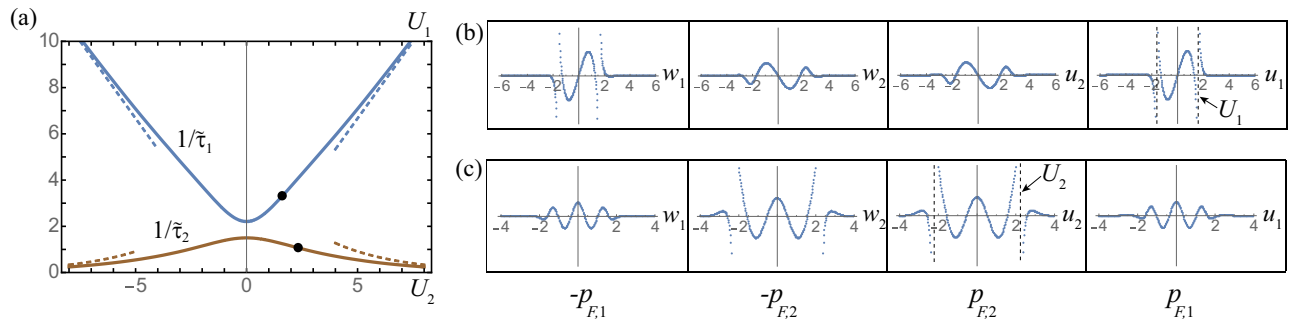}
\caption{Relaxation modes display singularities either in the first or second channel. (a) The relaxation rates $1/\tilde{\tau}_1(U_1)$ and $1/\tilde{\tau}_2(U_2)$ for modes with singularities in band 1 at $u_1 = U_1$ and in band 2 at $u_2 = U_2$, respectively. The dashed lines correspond to the decay rates (\ref{eq:tauk1form}) and (\ref{eq:tauk2form}) of high energy quasiparticles. The black dots indicate the rates of the modes shown in (b) and (c). (b) A typical mode with singularities at $u_1 = \pm U_1$ and $w_1 = \pm U_1$, indicated by vertical asymptotes, whose eigenvalue is $1/\tilde{\tau}_1(U_1)$. (c) Another typical mode with singularities at $u_2 = \pm U_2$ and $w_2 = \pm U_2$ and an eigenvalue $1/\tilde{\tau}_2(U_2)$. The four panels give the eigenfunction near the Fermi points $-p_{F,1}$, $-p_{F,2}$, $p_{F,2}$, and $p_{F,1}$.}
\label{fig:nonzeromodes}
\end{figure*}
\end{center}

\begin{center}
\begin{figure*}
\includegraphics[width = 12cm]{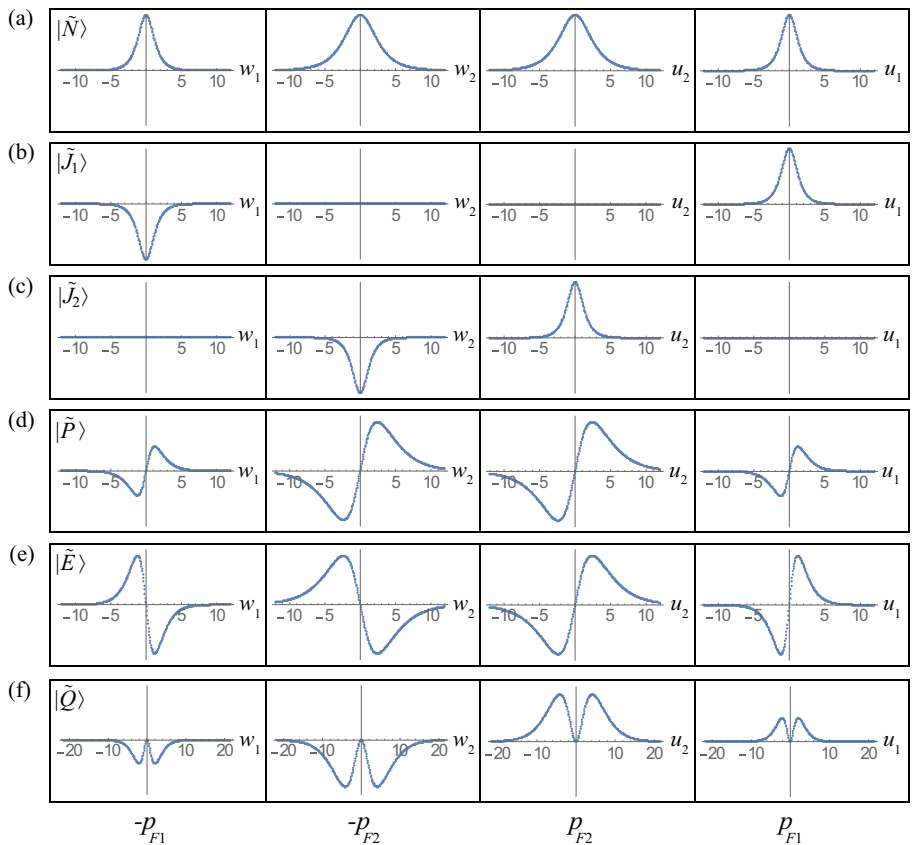}
\caption{Six zero modes of the collision integral for system with two channels and a linear fermionic spectrum. The four horizontal panels show plots of the eigenfunction near the Fermi points $-p_{F,1}$, $-p_{F,2}$, $p_{F,2}$, and $p_{F,1}$. The mode label is given in the left-most panel.}
\label{fig:zeromodes}
\end{figure*}
\end{center}

\subsection{Numerical Solution}

We now present a numerical solution of the eigenvalue problem (\ref{eq:eigprob}) for the case of $l = 2$ channels with a linearized spectrum. The numerical solution requires recasting the integral equation~(\ref{eq:xdotki}) in terms of dimensionless variables. To this end, we introduce
\begin{equation}
u_i = \frac{v_1 k_i}{2 T}, \quad  w_i = \frac{v_1 q_i}{2T}, \quad \tilde{t} = \frac{t}{t_0},
\label{eq:dimless}
\end{equation}
where
\begin{equation}
\frac{1}{t_0} = \frac{ |\lambda|^2 T}{8 \pi \hbar^3 v_1 v_2}.
\label{eq:t0}
\end{equation}
The dimensionless form of Eq.~(\ref{eq:xdotki}) for the two channel case is given by
\begin{eqnarray}
\nonumber
\frac{d \tilde{x}_{u_1}}{d \tilde{t}} &=& - \frac{1}{\tilde{\tau}_1(u_1)} \tilde{x}_{u_1} - \int \! du_2 \tilde{\Gamma}_{12}^{++}(u_1,u_2) \tilde{x}_{u_2} \nonumber \\ &&  - \int  dw_{1} \tilde{\Gamma}_{11}^{+-}(u_1,w_1) \tilde{x}_{w_1} \nonumber \\
&&  - \int dw_{2} \tilde{\Gamma}_{12}^{+-}(u_1,w_2) \tilde{x}_{w_2}.
\label{eq:xdotui}
\end{eqnarray}
Similarly, the three quantities $d \tilde{x}_{u_2}/d \tilde{t}$ and $d \tilde{x}_{w_i} / d \tilde{t}$ for $i = 1,2$ can be expressed in terms of $1/\tilde{\tau}_i(u_i)$ and $\tilde{\Gamma}_{ij}^{rs}$ which are defined in direct analogy with Eq.~(\ref{eq:xdotui}). The explicit forms for $1/\tilde{\tau}_i(u_i)$ and the various $\tilde{\Gamma}_{ii'}^{rr'}$ are given in Appendix B. The functions $1/\tilde{\tau}_1(u_1)$ and $1/\tilde{\tau}_2(u_2)$ play an important role in the analysis of the spectrum of the collision integral. The eigenmodes $| \tilde{X}_j \rangle$ of the operator $\tilde{\Gamma}$ defined by the right hand side of Eq.~(\ref{eq:xdotui}) are the solutions of the dimensionless eigenvalue problem
\begin{equation}
\tilde{\Gamma} | \tilde{X}_j \rangle = \tilde{\gamma}_j | \tilde{X}_j \rangle.
\label{eq:eigprobdim}
\end{equation}
The eigenvalues in Eq.~(\ref{eq:eigprob}) are given by $\gamma_j = \tilde{\gamma_j} / t_0 $.

We have obtained a numerical solution of the eigenvalue problem~(\ref{eq:eigprobdim}) for several values of $v_2/v_1$. The solutions include both zero modes and modes with finite relaxation rates. The latter are responsible for returning the system to equilibrium. The relaxation spectrum is shown in Fig.~\ref{fig:nonzeromodes}(a) and two typical modes are shown in Figs.~\ref{fig:nonzeromodes}(b) and (c). A salient characteristic of these modes is the presence of four singularities in the same band, such as at $u_1 = w_1 = \pm U_1$ for band 1 or $u_2 = w_2 = \pm U_2$ for band 2. Moreover, the eigenvalues of the modes are given by $1/\tilde{\tau}_1(U_1)$ or $1/\tilde{\tau}_2(U_2)$, depending on whether the singularity resides in band 1 or band 2. The full spectrum of relaxation rates corresponds to all possible values of $U_1$ and $U_2$. As shown in Fig.~\ref{fig:nonzeromodes}(a), the functions $1/\tilde{\tau}_1(U_1)$ and $1/\tilde{\tau}_2(U_2)$ have non-overlapping ranges and thus the spectrum of relaxation rates possesses a gap but is otherwise continuous. As depicted in Fig.~\ref{fig:nonzeromodes}(a), at $|U_{1/2}| \to \infty$ the relaxation rates $1/\tilde{\tau}_1(U_1)$ and $1/\tilde{\tau}_2(U_2)$ reproduce the decay rates~(\ref{eq:tauk1form}) and (\ref{eq:tauk2form}) of high energy quasiparticles.

To examine why the functions $1/\tilde{\tau}_1(u_1)$ and $1/\tilde{\tau}_2(u_2)$ control the spectrum of non-zero eigenvalues, we consider the mode shown in Fig.~\ref{fig:nonzeromodes}(b). First, suppose that the collision integral only possesses on-diagonal terms, such as $-x_{u_1}/\tilde{\tau}_1(u_1)$ on the right hand side of Eq.~(\ref{eq:xdotui}). In that case, the vector
\begin{equation}
| U_1 \rangle = \left(
                 \begin{array}{c}
                    \delta(u_1 - U_1)  \\
                    0 \\
                    0 \\
                    0 \\
                 \end{array}
               \right),
\label{eq:U0}
\end{equation}
would be an eigenmode of the operator $\tilde{\Gamma}$ with eigenvalue $-1/\tilde{\tau}_1(U_1)$. Once the off-diagonal terms $\tilde{\Gamma}_{ii'}^{rr'}$ are included, the eigenfunctions become non-zero for points away from $U_1$, however a singularity of the form $1/(u_1 - U_1)$ remains and the eigenvalue of the mode is still given by $-1/\tilde{\tau}_1(U_1)$. The mode shown in Fig.~\ref{fig:nonzeromodes}(b) illustrates these generic properties shared by all modes with non-zero eigenvalues. In contrast to $|U_1\rangle$ [defined in Eq.~(\ref{eq:U0})] which displays one singularity, the eigenfunctions of $\tilde{\Gamma}$ exhibit four singularities at the positions $u_1 = \pm U_1$ and $w_1 = \pm U_1$. This is consistent with the fact that $\tilde{\Gamma}$ respects both time-reversal and particle-hole symmetries, the latter arising from the linearized spectrum. The same reasoning naturally holds for modes with singularities in band 2, see Fig. 2(c).

In addition to modes with finite eigenvalues, the spectrum also contains six zero modes. The numerically obtained zero modes are found to be linear combinations of the modes shown in Fig.~\ref{fig:zeromodes}. Five of these modes have the form anticipated in Sec.~\ref{sec:zero} when cast in terms of the dimensionless variables $u_i$ and $w_i$. For instance, the zero mode $| N \rangle$ takes the dimensionless form
\begin{equation}
| \tilde{N} \rangle = \left(
                 \begin{array}{c}
                  \tilde{g}_{u_1} \\
                  \tilde{g}_{u_2} \\
                  \tilde{g}_{w_2} \\
                  \tilde{g}_{w_1} \\
                 \end{array}
               \right),
\label{eq:tildeN}
\end{equation}
where
\begin{equation}
\tilde{g}_{u_1} = \frac{1}{2 \cosh u_1}, \quad \tilde{g}_{u_2} = \frac{1}{2 \cosh \left(\frac{v_2}{v_1} u_2 \right)},
\end{equation}
and $\tilde{g}_{w_i}$ is given by replacing $u_i$ with $w_i$ in these expressions. The mode $|\tilde{N}\rangle$ is shown in Fig. 3(a). The modes $|\tilde{E}\rangle$, $|\tilde{P}\rangle$, $|\tilde{J}_1\rangle$, and $|\tilde{J}_2\rangle$, correspond to those given in Eqs.~(\ref{eq:Ezero}), (\ref{eq:Pzero}), (\ref{eq:J1}), and (\ref{eq:J2}), respectively. These modes are displayed in Figs.~\ref{fig:zeromodes} (b)-(e). An unexpected finding is that the spectrum possesses an additional zero mode, shown in Fig.~\ref{fig:zeromodes} (f). This mode is
\begin{equation}
| \tilde{Q} \rangle = \left(
                 \begin{array}{c}
 \, \, \, \, v_1 u_1^2 \tilde{g}_{u_1} \\
 \, \, \, \, v_2 u_2^2 \tilde{g}_{u_2} \\
           - v_2 w_2^2 \tilde{g}_{w_2} \\
           - v_1 w_1^2 \tilde{g}_{w_1} \\
                 \end{array}
               \right).
\label{eq:qtilde}
\end{equation}
The origin of this sixth zero mode is discussed in Sec.~\ref{sec:Q}.

\section{Bulk Viscosity}

\label{sec:viscosity}

While the average lifetimes of quasiparticles can be measured directly in specialized experiments~\cite{barak}, macroscopic phenomenona such as transport involve the lifetimes of all quasiparticles in aggregate. In this section we perform an explicit calculation of the bulk viscosity based directly on the spectrum of the collision integral. We find that the bulk viscosity can be expressed as a weighted average of the relaxation times associated with all the eigenmodes of the linearized collision integral with finite eigenvalues and the appropriate symmetries.

A gas moving with a velocity $u$ can be in equilibrium. However, if $u$ varies in position then the gas is undergoing expansion or compression. Such a system is out of equilibrium. As it relaxes, entropy is generated at a rate proportional to the bulk viscosity and the square of the gradient of the velocity, $\partial_x u$~\cite{landau-fluids}.

We considered the entropy generated in a multi-channel Fermi gas subject to a non-uniform velocity $u$ in Ref.~\cite{us}. This allowed us to express $\zeta$ in terms of the linearized collision integral describing the relaxation properties of the system. Adapting Eq.~(11) of Ref.~\cite{us} to our dimensionless variables, we find
\begin{equation}
\zeta = \frac{32 \sigma \hbar^2 v_2 \Delta^2}{T | \lambda |^2}.
\label{eq:zeta}
\end{equation}
with
\begin{equation}
\sigma = \langle \phi | \tilde{\Gamma}^{-1} | \phi \rangle.
\label{eq:sigma}
\end{equation}
Here, $| \phi \rangle$ describes the correction to the equilibrium distribution arising from non-zero $\partial_x u$. It is given by
\begin{equation}
| \phi \rangle = \frac{1}{v_1 + v_2}   \left(
                 \begin{array}{c}
     \, \, \, \, v_1 \tilde{g}_{u_1} \\
                -v_2 \tilde{g}_{u_2} \\
                -v_2 \tilde{g}_{w_2} \\
     \, \, \, \, v_1 \tilde{g}_{w_1} \\
                 \end{array}
               \right).
\label{eq:X}
\end{equation}
In Eq.~(\ref{eq:sigma}), we took
\begin{equation}
\langle \psi | \phi \rangle = \sum_i \left(  \int_{-\infty}^\infty d u_i \, \psi_{u_i} \phi_{u_i} + \int_{-\infty}^\infty d w_i \, \psi_{w_i} \phi_{w_i} \right)
\label{eq:inner}
\end{equation}
as the definition of the inner product of $| \psi \rangle$ and $| \phi \rangle$.

In order for Eq.~(\ref{eq:sigma}) to be well defined, $| \phi \rangle$ must be orthogonal to all the zero modes of the collision integral. We observe that $| \phi\rangle$ is symmetric under the transformation $u_i \leftrightarrow -w_i$. Since $|\tilde{J}_1\rangle$, $|\tilde{J}_2\rangle$, $|\tilde{P}\rangle$, and $|\tilde{Q
}\rangle$ are antisymmetric under this transformation as is clear from Fig.~3, these four modes are orthogonal to $|\phi \rangle$. To see that $|\phi\rangle$ is orthogonal to $|\tilde{E}\rangle$, we note that $|\phi \rangle$ is symmetric under $u_i \rightarrow -u_i$, $w_i \rightarrow -w_i$, whereas $|\tilde{E}\rangle$ is antisymmetric. Using Eqs.~(\ref{eq:tildeN}),~(\ref{eq:X}) and (\ref{eq:inner}), it can be verified that $\langle \tilde{N} | \phi \rangle = 0$.

The evaluation of the matrix element (\ref{eq:sigma}) requires inverting the linearized collision integral. This is most readily accomplished by writing $|\phi\rangle$ in terms of the normalized eigenmodes $|\tilde{X}_i\rangle$ of $\tilde{\Gamma}$. Expanding $|\phi\rangle$ in the basis $| \tilde{X}_i \rangle$, we obtain
\begin{equation}
\sigma = \sideset{}{'}\sum_j |\langle \tilde{X}_j | \phi \rangle|^2 \frac{1}{\tilde{\gamma}_j},
\label{eq:matrixelement}
\end{equation}
where the prime indicates that the sum excludes zero modes. Using the numerical solution of the eigenvalue problem (\ref{eq:eigprobdim}), we have obtained $\sigma$ for several values of $v_2/v_1$ as given in Table I.

\begin{table}
\caption{The quantity $\sigma$ for several values of $v_2/v_1$.}
\begin{center}
\begin{tabular}{c|c}
  \hline
  \hline
  $v_2/v_1$ & $\sigma$  \\
  \hline
  0.2 & 0.27 \\
  0.4 & 0.19 \\
  0.6 & 0.16  \\
  0.8 & 0.14  \\
  \hline
  \hline
\end{tabular}
\end{center}
\end{table}

The expression for the bulk viscosity (\ref{eq:zeta}) and the values of $\sigma$ in Table I complete our evaluation of $\zeta$ in terms of the microscopic parameters of the system. In contrast to the careful treatment of the relaxation spectrum presented here, in Ref.~\cite{us} $\zeta$ was crudely estimated using the relaxation-time approximation.  In that approach, the non-zero values of $\tilde{\gamma}_j$ are assumed to be equal to the same constant. This corresponds to the assumption that the relaxation times of all the eigenmodes are equal. The present work illustrates the shortcomings of this approximation and the necessity of taking into account the full spectrum of the linearized collision integral. As Eq.~(\ref{eq:matrixelement}) demonstrates, the bulk viscosity $\zeta$ (\ref{eq:zeta}) depends on the properly weighted average of the relaxation times of the eigenmodes of the linearized collision integral. The solution of the eigenvalue problem (\ref{eq:eigprobdim}) allows us to determine the manner in which $\zeta$ varies with $v_2/v_1$.

\section{An Additional Conserved Quantity}
\label{sec:Q}

A salient feature of the spectrum studied in Sec.~\ref{sec:spectrum} was the presence of an additional, unexpected zero mode $| \tilde{Q} \rangle$ given in Eq.~(\ref{eq:qtilde}). In this section we demonstrate that this zero mode, like the other five zero modes, is associated with a conserved quantity of the collision integral $\hat{\Gamma}$.

We now identify this conserved quantity, which we will denote by $Q$. In terms of the momenta $k_i$ and $q_i$, the eigenmode takes the form
\begin{equation}
| Q \rangle = \left(
                 \begin{array}{r}
         \, \, \, \, g_{k_1} v_1 k_1^2 \\
         \, \, \, \, g_{k_2} v_2 k_2^2 \\
                   - g_{q_2} v_2 q_2^2 \\
                   - g_{q_1} v_1 q_1^2 \\
                 \end{array}
               \right).
\label{eq:QmodeA}
\end{equation}
We may obtain $| \dot{Q} \rangle$ by substituting the components of $|Q\rangle$ into Eq.~(\ref{eq:fermi2}), i.e., taking $x_{k_1} = g_{k_1} v_1 k_1^2$, $x_{k_2} = g_{k_2} v_2 k_2^2$, etc. In particular, the terms in parentheses in Eq.~(\ref{eq:fermi2}) contribute a factor
\begin{equation}
v_1 k_1^2 - v_1 q_1^2 - v_2 k_2^2 + v_2 q_2^2
\label{eq:Qcons}
\end{equation}
to each of the terms in the sum. In analogy with the other five zero modes, we expect the expression (\ref{eq:Qcons}) to vanish, which is tantamount to the statement that the quantity
\begin{equation}
Q = Q_{1R} + Q_{2R} + Q_{2L} + Q_{1L},
\label{eq:Qcons2}
\end{equation}
where
\begin{equation}
Q_{iR} =  \sum_{k_i} v_i k_i^2  n_{k_i}, \ \ Q_{iL} = - \sum_{q_i} v_i q_i^2 n_{q_i}
\label{eq:Qcons3}
\end{equation}
is conserved by $\hat{\Gamma}$. That $Q$ is conserved by the processes shown in Fig.~1 follows from momentum and energy conservation. Conservation of momentum requires that
\begin{equation}
k_1 + q_1 = k_2 + q_2.
\label{eq:momen}
\end{equation}
For the linearized spectrum (\ref{eq:linearspec}), energy conservation gives
\begin{equation}
v_1 k_1 - v_1 q_1 = v_2 k_2 - v_2 q_2.
\label{eq:energy}
\end{equation}
The product of Eqs.~(\ref{eq:momen}) and (\ref{eq:energy}) immediately shows that the expression~(\ref{eq:Qcons}) is equal to zero and thus $Q$ given by Eqs.~(\ref{eq:Qcons2}) and (\ref{eq:Qcons3}) is a conserved quantity.

Ostensibly, the conservation of $Q$ hinges on the linearity of the fermionic spectrum. It is thus natural to ask what effect band curvature has on this mode. To address this question, we repeat the above analysis with a massive spectrum given by Eqs.~(\ref{eq:epsilon3A}) and (\ref{eq:epsilon3B}). Conservation of energy then gives
\begin{equation}
v_1 (k_1 - q_1) + \frac{k_1^2+q_1^2}{2m} = v_2 (k_2 - q_2) + \frac{k_2^2+q_2^2}{2m},\quad \label{eq:massive}
\end{equation}
where again $k_1, q_1,...$ are measured from the Fermi point. Following the procedure in the previous paragraph, we take the product of Eqs.~(\ref{eq:momen}) and (\ref{eq:massive}), which gives
\begin{eqnarray}
 v_1 k_1^2 &+& \frac{k_1^3}{2m} - v_1 q_1^2 + \frac{q_1^3}{2m} + \frac{k_1^2 q_1 + q_1^2 k_1}{2m}  \nonumber \\ &=& v_2 k_2^2 + \frac{k_2^3}{2m} - v_2 q_2^2 + \frac{q_2^3}{2m} + \frac{k_2^2 q_2 + q_2^2 k_2}{2m}. \label{eq:ep}
\end{eqnarray}
The cross terms in this equation prevent any straightforward interpretation of it as the expression of a conserved quantity. However, such cross terms can be eliminated. To do so, we first obtain
\begin{equation}
\frac{1}{6m} \left( k_1 + q_1 \right)^3 = \frac{1}{6m} \left( k_2 + q_2 \right)^3,
\label{eq:cube}
\end{equation}
which follows directly from Eq.~(\ref{eq:momen}). Subtracting Eq.~(\ref{eq:cube}) from Eq.~(\ref{eq:ep}), we obtain
\begin{equation}
v_1 k_1^2 + \frac{k_1^3}{3m} - v_1 q_1^2 + \frac{q_1^3}{3m} = v_2 k_2^2 + \frac{k_2^3}{3m} - v_2 q_2^2 + \frac{q_2^3}{3m}.
\end{equation}
This equation can be interpreted as the conservation of the quantity $Q$ generalized to the massive case, i.e., Eq.~(\ref{eq:Qcons2}) holds, where now
\begin{eqnarray}
Q_{iR} &=&  \sum_{k_i} \left( v_i k_i^2 + \frac{k_i^3}{3m} \right) n_{k_i}, \label{eq:Qi1} \\ Q_{iL} &=& \sum_{q_i} \left( - v_i q_i^2 + \frac{q_i^3}{3m} \right) n_{q_i}. \label{eq:Qi2}
\label{eq:Qconsm}
\end{eqnarray}
The corresponding zero mode is thus
\begin{equation}
|Q \rangle = \left(
                 \begin{array}{r}
       \, \, \, \,  \left( v_1 k_1^2 + \frac{k_1^3}{3m}  \right) g_{k_1}  \\
       \, \, \, \,  \left( v_2 k_2^2 + \frac{k_2^3}{3m}  \right) g_{k_2}  \\
                    \left( -v_2 q_2^2 + \frac{q_2^3}{3m} \right) g_{q_2}  \\
                    \left( -v_1 q_1^2 + \frac{q_1^3}{3m} \right) g_{q_1}  \\
                 \end{array}
               \right).
\label{eq:Qprime}
\end{equation}
This argument readily generalizes to any number of channels $l$.

As seen above, $|Q\rangle$ remains a zero mode even for a spectrum with a finite mass $m$. We now discuss the case in which which the masses for each band are distinct, i.e, the multichannel spectrum is given by
\begin{eqnarray}
\varepsilon_{k_i} &=& v_i k_i + k_i^2/2m_i, \label{eq:diffmass1} \\
\varepsilon_{q_i} &=& - v_i q_i + q_i^2/2m_i, \label{eq:diffmass2}
\end{eqnarray}
where $m_i \neq m_j$ for $i \neq j$. An attempt to repeat the derivation of $Q$ leading to Eqs.~(\ref{eq:Qi1}) and (\ref{eq:Qi2}) is only partially successful. Elimination of the cubic cross terms, analogous to those in Eq.~(\ref{eq:ep}) generates new quartic terms. This procedure may be repeated, but at a certain order it becomes impossible to eliminate all of the cross terms. Thus, we anticipate that the mode corresponding to $|Q\rangle$ will acquire a finite relaxation rate $1/\tau_Q$. The order in momentum at which it becomes impossible to eliminate cross terms is directly related to the scaling of $1/\tau_Q$ with temperature. We find two cases, depending on the number of channels $l$,
\begin{equation}
\frac{1}{\tau_Q} \propto
\begin{cases} \, \, T^7 \, \, \, \mbox{for } l = 2, \\
              \, \, T^3 \, \, \, \mbox{for } l \geq 3. \\
\end{cases}
\label{eq:Qrates}
\end{equation}
This scaling is confirmed by numerical simulations. The mode $|Q\rangle$ is thus a slow mode---its relaxation rate is small compared to the typical rate $1/t_0 \propto T$ of the other modes.

\section{Discussion and Conclusions}
\label{sec:conclusion}

In this paper we have studied the relaxation of multi-channel Fermi gases. We have presented a detailed study of the linearized collision integral describing the scattering processes that dominate the equilibration properties at low temperatures, see Fig.~1. Focusing on the two-channel case, we found that the relaxation spectrum contains six zero modes as well as two branches of continuous spectrum comprised of modes with finite relaxation rates. A salient feature of the latter are eigenfunctions with singularities present in exactly one of the bands. The relaxation rates of the modes are given by $1/t_0\tilde{\tau}_1(U)$ or $1/t_0\tilde{\tau}_2(U)$, where $U = v_1 k / 2T$ and $k$ is the momentum at which the singularity occurs, measured from the nearest Fermi point. In the limit of energies large compared to the temperature, the relaxation rates of these modes correspond to the decay rates of single-particle excitations.

In addition to modes with finite relaxation rates, there are six zero modes, each of which corresponds to a conserved quantity. Three of these modes stem from fundamental conservation laws, to wit, particle number, momentum, and energy conservation. The other three zero modes derive from conservation laws specific to the scattering processes considered here. The conservation of $J_1$ and $J_2$ is not fundamental; indeed, it is violated by three-particle processes involving holes near the bottoms of the bands~\cite{lunde,holes}. These and other processes, not included in our treatment, endow $|J_1\rangle$ and $|J_2\rangle$ with finite but exponentially small relaxation rates $\tau_1^{-1}$ and $\tau_2^{-1}$, respectively. The criteria for $Q$ to be conserved are even more stringent, requiring both the particular type of collision~(\ref{eq:rxn}) as well as the form (\ref{eq:quad}) of the energy spectrum. Violation of these conditions results in a nonvanishing relaxation rate $\tau_Q^{-1}$ for the mode $|Q\rangle$. For example, if the masses for each channel are distinct, as in Eqs.~(\ref{eq:diffmass1}) and (\ref{eq:diffmass2}), then $|Q\rangle$ acquires a finite relaxation rate (\ref{eq:Qrates}) which is large compared with the exponentially small rates $\tau_1^{-1}$ and $\tau_2^{-1}$ but small compared with those of all the other modes with finite relaxation rates.

As an application of our analysis of the relaxation spectrum, we calculated the bulk viscosity $\zeta$ of a two-channel Fermi system. Importantly, $\zeta$ is not affected by the modes $|J_1 \rangle$, $|J_2\rangle$, and $|Q\rangle$. This follows from the fact that these modes are odd under inversion, $x \to -x$, while the bulk viscosity quantifies the response of the system to the gradient of the velocity $\partial_x u$, which is even. In previous work, $\zeta$ was estimated using the relaxation time approximation~\cite{us}. This approach predicted that $\zeta$ diverges as $1/T$ in the limit $T \to 0$, in contrast to the behavior of the bulk viscosity of a Fermi liquid in three dimensions, which vanishes in the same limit~\cite{sykes}. The divergence of $\zeta$ highlights the central role that the bulk viscosity likely plays in the non-equilibrium properties of 1D Fermi systems. While confirming this scaling behavior, the present work goes far beyond the relaxation time approximation and gives a detailed calculation of $\zeta$. This allowed for a precise determination of the dependence of the bulk viscosity on $v_2/v_1$ which, in turn, can be cast in terms of other microscopic parameters characterizing the system.

Although the symmetry of the modes $|J_1\rangle$, $|J_2\rangle$, and $|Q\rangle$ prevents them from contributing to the bulk viscosity, they do influence the thermal conductivity $\kappa$.  Indeed, the latter quantifies the response of the system to a temperature gradient $\partial_x T$, which is odd under inversion. Since in general the thermal conductivity is proportional to the relaxation time, we expect that in the static regime it will be controlled by the exponentially long times $\tau_1$ and $\tau_2$. In contrast, the frequency dependent thermal conductivity $\kappa(\omega)$ would reveal the full hierarchy of relaxation rates, as has been recently demonstrated in the single channel case~\cite{mirlin,zoran}.

Recent work on the emergence of two-fluid hydrodynamics in single channel systems~\cite{secondsound,attenuation} highlights an important principle: at frequencies far in excess of a relaxation rate of a discrete mode, the system responds as if that mode is a zero mode with an associated conserved quantity. For frequencies $\omega \ll \tau_1^{-1}, \tau_2^{-1}$, only the fundamental quantities $N$, $P$, and $E$ are conserved, leading to the standard hydrodynamic equations~\cite{landau-fluids}. At a frequency $\omega$ in the range $\tau_1^{-1}, \tau_2^{-1} \ll \omega \ll \tau_Q^{-1}$, two additional quantities, $J_1$, and $J_2$, are conserved. For $\tau_Q^{-1} \ll \omega \ll t_0^{-1}$, a sixth quantity, derived from $|Q\rangle$, would also be effectively conserved. Each new conserved quantity would give rise to an additional hydrodynamic equation. A natural avenue for future study is the investigation of the physics engendered by these atypical conservation laws.

\begin{acknowledgements}
The authors would like to thank A. V. Andreev for discussions. W.~D. acknowledges support from JQI and IREAP. Work by K.~A.~M. was supported by the U.S. Department of Energy, Office of Science, Materials Sciences and Engineering Division.
\end{acknowledgements}

\begin{appendix}

\section{Collision kernel for the two channel case}

In this Appendix, we give the explicit form of the functions appearing in Eq.~(\ref{eq:xdotki}) and specify the procedure for obtaining the functions which fully describe the linearized collision integral.
\begin{widetext}
The diagonal term in Eq.~(\ref{eq:xdotki}), obtained from Eq.~(\ref{eq:fermi2}), is given by
\begin{equation}
\frac{1}{\tau_1(k_1)} =
\frac{|\lambda|^2}{16 \pi \hbar^3 v_2} \cosh \left( \frac{v_1 k_1}{2T} \right) \int_{-\infty}^{\infty} \frac{dq_1}{\cosh \left( \frac{v_1 q_1}{2T} \right) \cosh \left( \frac{(v_1-v_2) k_1 - (v_1 + v_2) q_1}{4T} \right) \cosh \left( \frac{(v_1-v_2)q_1 - (v_1 + v_2) k_1}{4T} \right)}.
\label{eq:tau1}
\end{equation}
The off-diagonal terms are given by the three functions
\begin{equation}
\Gamma_{12}^{++}(k_1,k_2) = -\frac{|\lambda|^2}{8 \pi \hbar^3 \left|v_1 - v_2 \right|} \frac{1}{\cosh \left( \frac{v_1+v_2}{v_1-v_2}\frac{v_1 k_1}{2T} - \frac{v_1 v_2}{v_1-v_2} \frac{k_2}{T} \right) \cosh \left( \frac{v_1+v_2}{v_1-v_2}\frac{v_2 k_2}{2T} - \frac{v_1 v_2}{v_1-v_2} \frac{k_1}{T} \right)},
\label{eq:Gamma12pp}
\end{equation}
as well as
\begin{equation}
\Gamma_{11}^{+-}(k_1,q_1) = \frac{|\lambda|^2}{16 \pi \hbar^3 v_2} \frac{1}{\cosh \left( \frac{(v_1-v_2) k_1 - (v_1 + v_2) q_1}{4T} \right) \cosh \left( \frac{(v_1-v_2)q_1 - (v_1 + v_2) k_1}{4T} \right)},
\end{equation}
and
\begin{equation}
\Gamma_{12}^{+-}(k_1,q_2) =  - \frac{|\lambda|^2}{8 \pi \hbar^3 (v_1 + v_2)} \frac{1}{\cosh \left( \frac{v_1-v_2}{v_1+v_2}\frac{v_1 k_1}{2T} + \frac{v_1 v_2}{v_1+v_2} \frac{q_2}{T} \right) \cosh \left( \frac{v_1-v_2}{v_1+v_2} \frac{v_2 q_2}{2T} - \frac{v_1 v_2}{v_1+v_2} \frac{k_1}{T} \right)}.
\label{eq:Gamma12pm}
\end{equation}

The linearized collision integral can be completely specified by the four quantities above as well as the analogous quantities $\tau_{i}$ and $\Gamma_{ii'}^{rr'}$ defined for $\dot{x}_{k_2}$, $\dot{x}_{q_1}$ and $\dot{x}_{q_2}$. In total, there are 16 such quantities: four for each of the equations that comprise the linearized collision integral. The other 12 quantities not explicitly given are readily obtained from either inversion symmetry or the permutation of the channel indices. For instance, $\Gamma_{12}^{-+}(q_1,k_2)$ can be obtained by taking $k_1 \rightarrow - q_1$ and $q_2 \rightarrow -k_2$ and inverting the $+/-$ indices in Eq.~(\ref{eq:Gamma12pm}), giving $\Gamma_{12}^{-+}(q_1,k_2) = \Gamma_{12}^{+-}(-q_1,-k_2)$. Permutation of the channel indices allows us to obtain $\Gamma_{21}^{++}(k_2,k_1)$ from Eq.~(\ref{eq:Gamma12pp}). Specifically, we take $v_1 \leftrightarrow v_2$ and $k_1 \leftrightarrow k_2$, which gives $\Gamma_{21}^{++}(k_2,k_1) = \Gamma_{12}^{++}(k_1,k_2)$.

\section{Dimensionless collision kernel}

In this Appendix, we give the quantities $\tilde{\tau}_i$ and $\tilde{\Gamma}_{ii'}^{rr'}$, which appear in Eq.~(\ref{eq:xdotui}). These may be obtained by expressing $\tau_i$ and $\Gamma_{ii'}^{rr'}$ (see Appendix A) in terms of the dimensionless variables introduced in Eq.~(\ref{eq:dimless}).

Expressing Eq.~(\ref{eq:tau1}) in terms of the dimensionless variables given in Eq.~(\ref{eq:dimless}), we obtain
\begin{equation}
\frac{1}{\tilde{\tau}_{1}(u_1)} = \cosh u_1 \int_{-\infty}^{\infty}  \frac{d w_1}{\cosh w_1 \cosh \left( \frac{1-\eta}{2} u_1 - \frac{1+\eta}{2} w_1 \right) \cosh \left( \frac{1-\eta}{2} w_1 - \frac{1+\eta}{2} u_1 \right)},
\label{eq:tu1}
\end{equation}
where $\eta = v_2/v_1$. As described at the end of Appendix A, $\tau_2(k_2)$ may be obtained from $\tau_1$ by interchanging the channel numbers $1$ and $2$ everywhere in Eq.~(\ref{eq:tau1}). This procedure gives
\begin{equation}
\frac{1}{\tilde{\tau}_2(u_2)} = \eta  \cosh \left( \eta u_2 \right) \int_{-\infty}^{\infty}  \frac{d w_2}{\cosh \left( \eta w_2 \right)  \cosh \left(  \frac{1+\eta}{2} u_2 +  \frac{1-\eta}{2} w_2 \right) \cosh \left(  \frac{1+\eta}{2} w_2 +  \frac{1-\eta}{2} u_2 \right)}.
\label{eq:tauu2}
\end{equation}

The three other terms in Eq.~(\ref{eq:xdotui}) are given by
\begin{equation}
\tilde{\Gamma}_{12}^{++}(u_1,u_2) = - \frac{2 \eta}{ \left| 1-\eta \right|}  \frac{1}{ \cosh \left( \frac{1+\eta}{1-\eta} u_1 - \frac{2\eta}{1-\eta} u_2 \right) \cosh \left( \frac{\eta+\eta^2}{1-\eta} u_2 - \frac{2 \eta}{1-\eta} u_1 \right)},
\label{eq:Gammat12pp}
\end{equation}
\begin{equation}
\tilde{\Gamma}_{11}^{+-}(u_1,w_1) =  \frac{1}{\cosh \left( \frac{1-\eta}{2} u_1 - \frac{1+\eta}{2} w_1 \right) \cosh \left( \frac{1+\eta}{2} u_1 - \frac{1-\eta}{2} w_1 \right)},
\end{equation}
and
\begin{equation}
\tilde{\Gamma}_{12}^{+-}(u_1,w_2) = -  \frac{2 \eta}{1 + \eta}  \frac{1}{\cosh \left( \frac{1-\eta}{1+\eta} u_1 + \frac{2 \eta}{1+\eta} w_2 \right) \cosh \left( \frac{\eta-\eta^2}{1+\eta}  w_2 - \frac{2\eta}{1+\eta} u_1  \right)}.
\label{eq:gammat12pm}
\end{equation}

An expression for $d \tilde{x}_{u_2} / d \tilde{t}$, analogous to Eq.~(\ref{eq:xdotui}), can be cast in terms of Eq.~(\ref{eq:tauu2}) and three additional quantities, again see the discussion at the end of Appendix A. These three quantities are
\begin{equation}
\tilde{\Gamma}_{22}^{+-}(u_2,w_2) =   \frac{\eta}{\cosh \left( \frac{1-\eta}{2} u_2 + \frac{1+\eta}{2} w_2 \right) \cosh \left( \frac{1+\eta}{2} u_2 + \frac{1-\eta}{2} w_2 \right)},
\end{equation}
\end{widetext}
$\tilde{\Gamma}_{21}^{++}(u_2,u_1)$, and $\tilde{\Gamma}_{21}^{+-}(u_2,w_1)$. The last two quantities are the dimensionless versions of $\Gamma_{21}^{++}(k_2,k_1)$ and $\Gamma_{21}^{+-}(k_2,q_1)$, respectively, which may be obtained by employing the prescription given at the end of Appendix A. This gives
\begin{eqnarray}
\tilde{\Gamma}_{21}^{++}(u_2,u_1) &=& \tilde{\Gamma}_{12}^{++}(u_1,u_2), \\
\tilde{\Gamma}^{+-}_{21}(u_2,w_1) &=& \tilde{\Gamma}_{12}^{+-}(w_1,u_2),
\label{eq:last}
\end{eqnarray}
and may thus be obtained from Eqs.~(\ref{eq:Gammat12pp}) and (\ref{eq:gammat12pm}). The eight quantities given in Eqs.~(\ref{eq:tu1})-(\ref{eq:last}) specify $d \tilde{x}_{u_1} / d \tilde{t}$ and $d \tilde{x}_{u_2} / d \tilde{t}$.

The eight remaining quantities appear in the expressions for $d \tilde{x}_{w_1} / d \tilde{t}$ and $d \tilde{x}_{w_2} / d \tilde{t}$ and are readily obtained from those already given here by invoking inversion symmetry. For instance, the expression $d \tilde{x}_{w_1}/d \tilde{t}$ contains the term $-\tilde{x}_{w_2}/\tilde{\tau}_1(w_1)$, which is analogous to the first term on the right hand side of Eq.~(\ref{eq:xdotui}). The fact that this term is given by Eq.~(\ref{eq:tu1}) follows directly from inversion symmetry. Similarly, $d \tilde{x}_{w_2}/d \tilde{t}$ contains a term $-\tilde{x}_{w_2}/\tilde{\tau}_2(w_2)$ where $\tilde{\tau}_2(w_2)$ is given by Eq.~(\ref{eq:tauu2}).
Expressions of the form $\tilde{\Gamma}^{-+}_{ij}(w_i, u_j)$ and $\tilde{\Gamma}^{--}_{ij}(w_i,w_j)$ may be expressed in terms of Eqs.~(\ref{eq:Gammat12pp})-(\ref{eq:last}) by again exploiting inversion symmetry. Inversion is enacted by negating the $+/-$ indices and both momenta. This gives
\begin{eqnarray}
\tilde{\Gamma}^{-+}_{ij}(w_i,u_j) &=& \tilde{\Gamma}^{+-}_{ij}(-w_i,-u_j), \\
\tilde{\Gamma}^{--}_{ij}(w_i,w_j) &=& \tilde{\Gamma}^{++}_{ij}(-w_i,-w_j).
\end{eqnarray}

\end{appendix}


\begin{thebibliography}{99}
\bibitem{footnote1} It should be emphasized that the activated form of Eq.~(\ref{eq:tauk2form}) is based exclusively on the effects of $\hat{V}_{1D}$ [Eq.~(\ref{eq:V1D})]. In fact, if three particle collisions are included, it is found that the scattering rate is proportional to a power of temperature, and thus three particle collisions would dominate the scattering rate for sufficiently low temperatures.
\end{thebibliography}
\end{document}